\documentclass[fleqn]{2023SCGE}
\usepackage{bm}
\usepackage{soul}
\usepackage{indentfirst}
\usepackage{amsmath}
\usepackage{graphicx}
\usepackage{float}
\usepackage{amssymb}
\usepackage{subfigure}
\usepackage{amssymb}
\usepackage{psfrag}
\usepackage{hyperref}
\usepackage{epstopdf}
\usepackage{tensor}
\usepackage{multirow}
\usepackage{comment}
\hypersetup{
  colorlinks=true,
  linkcolor=red,
  citecolor=blue,
}

\allowdisplaybreaks[1]
\begin{document}

\ensubject{subject}



\title{Reconciling Nonminimally Coupled  Higgs Inflation with ACT DR6 Observations through Reheating}
{Reconciling Nonminimally Coupled  Higgs Inflation with ACT  DR6 Observations through Reheating}

\author[1]{Lang~Liu}{}
\author[1]{Zhu Yi}{{yz@bnu.edu.cn}}
\author[2]{Yungui Gong}{{gongyungui@nbu.edu.cn}}

\address[1]{Department of Physics, Faculty of Arts and Sciences, Beijing Normal University, Zhuhai 519087, China}
\address[2]{Institute of Fundamental Physics and Quantum Technology, Department of Physics, School of Physical Science and Technology, \\ Ningbo University, Ningbo, Zhejiang 315211, China}

\abstract{The Higgs inflation model with nonminimal coupling, while disfavored by the 1$\sigma$ region of the latest Atacama Cosmology Telescope Data Release 6 (ACT DR6) observational data, can be reconciled with the ACT DR6 data by incorporating the effects of reheating. In this paper, we consider reheating with a constant equation of state $w_{re}$. For the strong coupling case $\xi=100$, we find that reconciling the model with both the ACT DR6 constraints and the minimum reheating temperature required for successful Big Bang Nucleosynthesis (BBN) demands $w_{re} \geq 0.92$. 
Specifically, the reheating $e$-folding number must be
$ N_{ re}= 28.3$ for $w_{ re} = 0.92$, and within $24.9 \leq N_{ re} \leq 27.2$ for $w_{ re} = 1$.  In the more general case without assuming the strong coupling limit, consistency with both ACT and BBN requires the nonminimal coupling to satisfy $\xi \geq 0.05$. 
Our findings suggest that by considering reheating, a wide range of inflationary models, such as $R^2$ inflation, hilltop inflation, E-model inflation, and T-model inflation, can also be made consistent with the ACT DR6 observational data.
}


\maketitle

\begin{multicols}{2}

\section{Introduction}
Inflation is a crucial component of modern cosmology that addresses several fundamental issues in the standard Big Bang model. By positing a period of rapid exponential expansion in the early universe, inflation provides a compelling explanation for the observed flatness, homogeneity, and isotropy of the cosmos on large scales. Moreover, inflation naturally accounts for the absence of magnetic monopoles and generates the primordial density fluctuations that serve as the seeds for 
\Authorfootnote
\noindent 
the formation of galaxies and large-scale structures \cite{Starobinsky:1980te, Guth:1980zm, Linde:1981mu, Albrecht:1982wi}.

Among the various inflationary models proposed, Higgs inflation \cite{Bezrukov:2007ep} stands out for its simplicity and connection to particle physics. 
In this scenario, the Higgs field, which is responsible for generating the masses of elementary particles in the Standard Model, also drives the inflationary expansion. The canonical Higgs inflation model, with the potential $V_h(\phi) = \lambda  \phi^4/4$, is ruled out by observational data due to its prediction of a large tensor-to-scalar ratio. However, introducing a nonminimal coupling term $\xi \phi^2 R$ allows for a reduction in the tensor-to-scalar ratio. In the strong coupling regime $\phi \gg 1/\sqrt{\xi}$, the scalar spectral index $n_s$ and the tensor-to-scalar ratio become $n_s - 1 = 2 / N$ and $r = 12 / N^2$, respectively \cite{Kaiser:1994vs, Bezrukov:2007ep}, where $N$ denotes the number of e-folds between horizon exit and the end of inflation. For $N = 60$, the model predicts $n_s = 0.9667$ and $r = 0.0033$, which are consistent with the {\it Planck} 2018 constraints \cite{Planck:2018jri}.

Recently, the Atacama Cosmology Telescope  Data Release 6 (ACT DR6) results \cite{Louis:2025tst, ACT:2025tim} have suggested a higher value of the scalar spectral index $n_s$ than previously reported by {\it Planck}. A combined analysis of {\it Planck} and ACT DR6 data (P-ACT) gives $n_s = 0.9709 \pm 0.0038$ \cite{Louis:2025tst}. When additional measurements from CMB lensing and Baryon Acoustic Oscillation (BAO) observations by the Dark Energy Spectroscopic Instrument (DESI) \cite{DESI:2024uvr, DESI:2024mwx} are included, the extended dataset (P-ACT-LB) yields $n_s = 0.9743 \pm 0.0034$ \cite{Louis:2025tst}, representing a $2\sigma$ deviation from the original {\it Planck} results. As a result, the Higgs inflation model with nonminimal coupling is disfavored under the updated P-ACT-LB constraints. Although the current discrepancy is at approximately the $1\sigma$ level, it is noteworthy that the inferred central value of $n_s$ has shifted systematically upward with the successive inclusion of independent datasets---from Planck alone ($n_s = 0.9649 \pm 0.0042$) to P-ACT ($n_s = 0.9709 \pm 0.0038$) and further to P-ACT-LB ($n_s = 0.9743 \pm 0.0034$). This persistent trend, driven by the addition of independent measurements, suggests that the shift may not be a mere statistical fluctuation. Moreover, next-generation CMB experiments such as CMB-S4~\cite{CMB-S4:2016ple}, LiteBIRD~\cite{Hazumi:2019lys}, and the Simons Observatory~\cite{SimonsObservatory:2018koc} are expected to achieve sensitivities of $\sigma(n_s) \sim 0.001$--$0.002$. If the central value of $n_s$ remains near its current level, the tension with the standard prediction $n_s \approx 0.967$ could grow to $3\sigma$ or beyond. It is therefore timely and important to investigate mechanisms that can reconcile the model with the data.

After inflation's accelerated expansion, the universe enters a cold, vacuum-like state dominated by the inflaton field.  To transition into the hot Big Bang phase, the energy stored in the inflaton must be transferred to, or become subdominant to, a thermal bath of particles, a process known as reheating, see \cite{Bassett:2005xm, Allahverdi:2010xz, Amin:2014eta, Lozanov:2019jxc} for a review. 
The resulting reheating temperature $T_{re}$ establishes the initial conditions for the universe's thermal evolution and affects the number of e-folds $N$ between horizon exit and the end of inflation \cite{Liddle:2003as}. This, in turn, impacts the inflationary predictions for observables such as the scalar spectral index $n_s$ and the tensor-to-scalar ratio $r$ \cite{Dai:2014jja, Cai:2015soa, Gong:2015qha, Dimopoulos:2017zvq, Fei:2017fub, Mishra:2021wkm, Chakraborty:2023ocr, Jeong:2023zrv,  Zhang:2024ldx,Ballardini:2024ado}.

By incorporating the effects of reheating, it may be possible to reconcile the Higgs inflation model with nonminimal coupling with the updated P-ACT-LB constraints. In this paper, we demonstrate that the tension between the Higgs inflation model with nonminimal coupling and the latest P-ACT-LB data can be alleviated when reheating effects are taken into account. Beyond addressing the current tension with ACT DR6 data, our analysis provides concrete quantitative constraints on the reheating parameters---including $w_{ re}$, $N_{ re}$, and $T_{ re}$---that are valuable for post-inflationary model building independently of the precise level of observational discrepancy. Moreover, the attractor relation $n_s - 1 \simeq -2/N_*$ is shared by a broad universality class of inflationary models, including $R^2$ inflation \cite{Starobinsky:1980te}, $\alpha$-attractors \cite{Kallosh:2013yoa}, and hilltop models \cite{Boubekeur:2005zm}, so that our results have implications extending well beyond Higgs inflation alone. Alternative approaches to explaining the P-ACT-LB results can be found in Refs.~\cite{Kallosh:2025rni,Kallosh:2025ijd, Gialamas:2025kef, Frob:2025sfq, Dioguardi:2025vci, Brahma:2025dio, Berera:2025vsu, Aoki:2025wld,Peng:2025bws, Dioguardi:2025mpp, Salvio:2025izr, Gao:2025onc,Gao:2025viy, Yi:2025dms,He:2025bli,Haque:2025uri, Zhu:2025twm}. Throughout this paper, we assume the equation of state during reheating $w$ is a constant.

The rest of this paper is organized as follows. In Section \ref{Higgs}, we provide a concise review of Higgs inflation with nonminimal coupling. We discuss the impact of reheating on the number of  $e$-folds $N_*$ in Section \ref{Reheating} and present our main results in Section \ref{results}. Finally, Section~\ref{Conclusion} summarizes our findings and discusses their implications.

\section{Higgs Inflation Model with Nonminimal Coupling} \label{Higgs}

In the unitary gauge, the action of Higgs inflation with the nonminimal coupling term $\xi \phi^2 R$ in the Jordan frame is given by:
\begin{equation}\label{act:higgs1}
    S= \int d^4x \sqrt{-\tilde{g}} \left[\frac{ \Omega(\phi)}{2}\tilde{R}-\frac{1}{2}(\tilde{\partial} \phi)^2 -V_h(\phi)\right],
\end{equation}
where   quantities with a tilde are defined in the Jordan frame,  $\Omega(\phi) = M_\mathrm{pl} ^2 +\xi \phi^2$,  $\xi$ represents a dimensionless coupling constant, $M_\mathrm{pl} = 1/\sqrt{8\pi G}$ is the reduced Planck mass. We set the units such that $M_\mathrm{pl}=c=1$, where $c$ is the speed of light. The Higgs potential is:
\begin{equation}\label{higgs:potential}
V_h(\phi)=\frac{\lambda}{4} \phi^4.
\end{equation}
By performing the conformal transformation
\begin{equation}
    g_{\mu\nu} =  \Omega(\phi) \tilde{g}_{\mu\nu},
\end{equation}
and introducing the canonically normalized field $\psi$  through
\begin{equation}
 d \psi ^ { 2 } = \left[ \frac { 3 } { 2 } \left( \frac {\Omega'(\phi) } { \Omega(\phi) } \right) ^ { 2 } + \frac { 1} { \Omega ( \phi ) } \right] d \phi ^ { 2 } , 
\end{equation}
 with $\Omega'(\phi) = d\Omega(\phi)/d\phi$, we can obtain the 
 action of the  Higgs inflation \eqref{act:higgs1} in the Einstein frame,
\begin{equation}
    S_E= \int d^4x \sqrt{-g} \left[\frac{ 1}{2}R-\frac{1}{2}(\partial \psi)^2 -U(\psi)\right],
\end{equation}
where  $g_{\mu\nu}$, $\psi$, and $R$ are the metric, field, and the Ricci scalar in the Einstein frame.  The Einstein-frame potential is given by
\begin{equation}
    U(\psi)=V_h(\phi)/\Omega^2(\phi),
\end{equation}
where  $U$ is understood as a function of the canonical field $\psi$, with  $\phi$ implicitly expressed in terms of $\psi$ through the field redefinition.
In the strong coupling regime where $\phi\gg 1/\sqrt{\xi}$, the Einstein-frame potential reduces to \cite{Bezrukov:2007ep}
\begin{equation}
\label{hissen1}
 U(\psi) = U_0\left[1 - \exp(-2\psi/\sqrt{6})\right]^{ 2}, 
\end{equation}
with $U_0 = \lambda/(4\xi^2)$. Under the slow-roll condition, the predictions of scalar spectral index $n_s$ and tensor to scalar ratio $r$ from the potential \eqref{hissen1} are \cite{Bezrukov:2007ep}
\begin{equation}\label{higgs:prediction}
    n_s = 1 -\frac{2}{N}, \quad r = \frac{12}{N^2},
\end{equation}
which are identical to those of $R^2$ inflation \cite{Starobinsky:1980te}. The Higgs inflation model is a special case of the more general nonminimal coupling inflation model with coupling term $\xi f(\phi) R$ and potential $V(\phi)= \lambda ^2 f(\phi)$. Under the strong coupling condition $3\xi^2 f'^2(\phi)/2\gg 1+\xi f(\phi)$, any function $f(\phi)$ yields predictions matching equation \eqref{higgs:prediction} \cite{Kallosh:2013tua}. For an $e$-folding number $N=60$, the predicted scalar spectral index is $n_s = 0.968$, which falls outside the 1$\sigma$ constraints of the P-ACT-LB data.

The Higgs inflation model with nonminimal coupling is disfavored by the P-ACT-LB data due to the large $e$-folding number required to match observations. However, if the $e$-folding number is increased to $N=75$, the predictions of the Higgs inflation model  with nonminimal coupling become $n_s=0.973$ and $r=0.0021$.  These values are consistent with the 1$\sigma$ constraints of the   dataset P-ACT-LB-BK18~\cite{ACT:2025tim}, which incorporates B-mode measurements from the BICEP and Keck telescopes at the South Pole (BK18) \cite{BICEP:2021xfz}.
The $e$-folding number depends on the epoch of reheating and the subsequent evolution of the universe. By taking into account the reheating phase, it may be possible to reconcile the Higgs inflation model with nonminimal coupling with the latest observational data.

The duration of the reheating phase and the post–inflationary evolution of the inflaton and radiation energy densities  can significantly impact the $e$-folding number. A prolonged reheating phase, or more generally, a stiff post-inflationary expansion history effectively increases the $e$-folding number, potentially bringing the predictions of the Higgs inflation model with nonminimal coupling  into agreement with the P-ACT-LB data. In the following section, we will explore the impact of reheating on the $e$-folding number.

\section{Reheating}
\label{Reheating}

The pivotal scale $k_*=0.05\rm Mpc^{-1}$  can be expressed in terms of the present Hubble horizon through the following relationship:
\begin{equation}
\label{kstar}
  \frac{k_\ast}{a_0 H_0}=\frac{a_\ast H_\ast}{a_0 H_0}=\frac{a_\ast}{a_{e}}\frac{a_{e}}{a_{re}}\frac{a_{re}}{a_0}\frac{H_\ast}{H_0}=e^{-N_\ast-N_{re}}\frac{a_{re}}{a_0}\frac{H_\ast}{H_0},
\end{equation}
where $a_*$, $a_e$, $a_{re}$, and $a_0=1$ denote the value of the scale factor at the horizon cross of $k_*$, end of inflation, end of reheating, and present, respectively.  $N_* =\ln (a_e/a_*)$ is the $e$-folding number between the horizon cross and end of inflation, $N_{re}=\ln(a_{re}/a_e)$ is the   $e$-folding number during reheating. 
We assume that the reheating epoch follows the inflation epoch immediately and the equation of state  $w_{re}$  during reheating is a constant, and the radiation epoch follows the reheating epoch immediately. This parametrization has a clear physical basis in the context of non-minimally coupled Higgs inflation. After the end of slow roll, the inflaton undergoes coherent oscillations around the minimum of its Einstein-frame potential. Near the minimum, the potential is well approximated by a power-law form $V(\phi) \propto \phi^n$. For such oscillations, the virial theorem gives a time-averaged equation of state $\langle w \rangle = (n-2)/(n+2)$. The constant $w_{ re}$ should therefore be understood as the time-averaged effective equation of state over the entire reheating epoch. This is a standard and widely adopted parametrization in the literature~~\cite{Dai:2014jja,Munoz:2014eqa,Cook:2015vqa}, whose principal advantage is that the shift in the number of e-folds $N_*$---and hence in the inflationary observables---depends on the integrated expansion history during reheating, which is governed by the average equation of state rather than by its detailed time dependence. Allowing $w_{ re}$ to vary within the range $-1/3 \leq w_{ re} \leq 1$ encompasses a broad spectrum of physically distinct reheating scenarios. By using the relation $\rho \propto a^{-3(1+w_{re})}$ during reheating,  we have
\begin{equation}
\label{reheq2}
N_{re}=\frac{1}{3(1+w_{re})}\ln\frac{\rho_{e}}{\rho_{re}},
\end{equation}
where $\rho_e$ and $\rho_{re}$  are the energy density at the end of inflation and reheating, respectively. We emphasize that the constant $w_{ re}$ parametrization is not merely a mathematical convenience but is in fact exact at the level of the integrated expansion history. During reheating, the energy density evolves as $d\ln\rho/dN = -3[1+w(N)]$, which upon integration yields
\begin{equation}
{\ln\!\left(\frac{\rho_{re}}{\rho_{e}}\right) 
= -3\bigl(1 + \bar{w}\bigr)\,N_{ re}\,,
\qquad
\bar{w} \equiv \frac{1}{N_{ re}}\int_0^{N_{ re}} w(N)\,dN\,,}
\end{equation}
where $\bar{w}$ is the $e$-fold--averaged equation of state. This identity holds for arbitrary time-dependent $w(N)$ without any approximation. Since the total expansion and energy-density ratio during reheating---which are the only quantities entering the derivation of $N_*$, $T_{ re}$, and $N_{ re}$---depend on $w(N)$ exclusively through $\bar{w}$, the parameter $w_{ re}$ in our analysis should be identified with $\bar{w}$. A time-varying $w(N)$ and a constant $w_{ re} = \bar{w}$ produce identically the same observational consequences. Our constraints on $w_{ re}$ are therefore constraints on the average thermodynamic history of reheating, valid regardless of the detailed time evolution of the equation of state. The relation of $\rho_{re}$ and the temperature $T_{re}$ at the end of reheating is 
\begin{equation}
\label{reheq1}
\rho_{re}=\frac{\pi^2}{30}g_{re}T^4_{re},
\end{equation}
where $g_{re}$ is the effective number of relativistic species at reheating. 
By using  the entropy conservation condition, we can obtain
\begin{equation}
\label{reheq3}
a_{re}^3 g_{s,re}T_{re}^3=a_0^3\left(2 T_0^3+6\times\frac78 T_{\nu0}^3\right),
\end{equation}
where $g_{s,re}$ is the effective number of relativistic species for entropy at reheating.  $T_0=2.725K$ and $T_{\nu0}=(4/11)^{1/3}T_0$ are the current cosmic microwave background temperature and neutrino temperature, respectively. 

By using the  above relation, we can obtain the exact $e$-folding number $N_*$ after the horizon cross and before the end of inflation
\cite{Dai:2014jja,Cook:2015vqa}
\begin{equation}
\begin{aligned}
 \label{Nre}
N_\ast=&-\frac{1-3w_{re}}{4}N_{re}-\ln\frac{\rho_{e}^{1/4}}{H_\ast}-\ln\frac{k_{\ast}}{a_0T_0}\\
&+\frac{1}{3}\ln\frac{43}{11g_{s,re}}+
  \frac14\ln\frac{\pi^2 g_{re}}{30}. 
\end{aligned}
\end{equation}
The temperature at the end of reheating can be expressed as 
\begin{equation}\label{eq:reh}
    T_{re}=\exp\left[-\frac{3N_{re}(1+w_{re})}{4}\right]\left[\frac{30\rho_{e}}{\pi^2 g_{re}}\right]^{1/4}.
\end{equation}
Given the logarithmic dependence of $N_{\rm re}$ and $T_{\rm re}$ on both $g_{\rm re}$ and $g_{s,\rm re}$, we approximate these parameters as $g_{\rm re}=g_{s,\rm re}=106.75$. Even allowing $g_{ re}$ to vary over its full Standard Model range from $106.75$ to $10.75$, the induced shift $|\Delta N_*|\simeq 0.19$ translates into $|\Delta n_s|\sim 10^{-4}$, which is negligible compared with the current observational uncertainty $\sigma(n_s)\simeq 0.004$. We conclude that the fixed $g_{re}$ approximation introduces a negligible systematic uncertainty compared to current and near-future observational errors.

For Higgs inflation with a nonminimal coupling, the scalar spectral index and the tensor-to-scalar ratio $r$ can be calculated from the Einstein-frame potential \eqref{hissen1} as
\begin{gather}\label{sl:ns}
    n_s -1 =-6\epsilon_V(\psi_*)+2\eta_V(\psi_*), \\
    \label{sl:r}
    r= 16 \epsilon_V(\psi_*),
\end{gather}
where the slow-roll parameters are defined by
\begin{equation}
    \epsilon_V(\psi) = \frac{1}{2}\left(\frac{U'(\psi)}{U(\psi)}\right)^2, \quad \eta_V(\psi) =\frac{U''(\psi)}{U(\psi)},
\end{equation}
and $U'(\psi)=dU(\psi)/d\psi$, $U''(\psi)=d^2U(\psi)/d\psi^2$. The field value $\psi_*$ at horizon crossing is related to the inflationary $e$-folding number $N_*$ through     
\begin{equation}
    N_* = \int_{\psi_e}^{\psi_*} \frac{U(\psi)}{U'(\psi)} d\psi= E(\psi_*)-E(\psi_e),
\end{equation}
with 
\begin{equation}
E(\psi) =\frac{1}{4}\left[3 \exp\left(\sqrt{\frac{2}{3}}\psi\right)-\sqrt{6} \psi\right].
\end{equation}
The field value $\psi_*$ at horizon crossing is then given by  
\begin{equation}\label{get_psi*}
\begin{aligned}
\psi_* &= -2\sqrt{\frac23}\bigg(N_*+E(\psi_e)\bigg)-\sqrt{\frac32}W_{-1}\bigg(-e^{-4[N_*+E(\psi_e)]/3}\bigg)\\
&\approx \sqrt{\frac{3}{2}}\ln\left(\frac{4N_*+4E(\psi_e)}{3}\right),
\end{aligned}
\end{equation}
where $W_{-1}$ denotes the $-1$ branch of the Lambert $W$ function. In obtaining the approximate expression, we have used the relation $W_{-1}(x)\approx \ln(-x)-\ln[-\ln(-x)]$ for $|x|\ll1$.

The field value at the end of inflation is determined by the condition  $\epsilon_V(\psi_e)=1$, which leads to $\psi_e = \sqrt{3/2}\ln(1+2/\sqrt{3})$.
The energy density of the inflation field at the end of inflation is
\begin{equation}
    \rho_e = \left(\frac{1}{2}\dot{\psi}^2 +U(\psi)\right)\bigg|_{\psi=\psi_e} 
    =1.5 U(\psi_e).
\end{equation} 

Using the approximate solution in Eq.~\eqref{get_psi*} and expanding the slow-roll parameters in the large-$N_*$ limit, we obtain
the scalar spectral index and tensor-to-scalar ratio to the leading order,
\begin{equation}\label{ns:ne}
    n_s -1 = - \frac{2}{N_* + n_e}, \quad r = \frac{12}{(N_*+n_e)^2},
\end{equation}
where  $n_e = E(\psi_e)-3/4\approx  0.3 $. 
The constant $n_e$ is fixed by the finite contribution of $\psi_e$ at the end of inflation together with the constant offset arising in the analytical derivation of $n_s$ and $r$.
The Hubble parameter at the horizon crossing is
given by 
 \begin{equation}
 \label{reheq4}
 H_* = \pi \sqrt{A_s r /2} \approx \sqrt{U(\psi_*)/3},
 \end{equation}
where $\ln (10^{10}A_s) = 3.06$ \cite{Louis:2025tst}.
Substituting the above relations into Eq. \eqref{Nre}, we obtain
\begin{equation}\label{N_Nre}
    N_* \approx  58  - \frac{1-3 w_{re}}{4}N_{re}.
\end{equation}
Combining it with Eqs. \eqref{ns:ne} and \eqref{eq:reh}, we have
\begin{equation}\label{ns:Nre}
  N_{re} \approx \frac{4[2- 58 (1-n_s)]}
  {(1-n_s)(3w_{re}-1)}, 
\end{equation}
\begin{equation}\label{ns:Tre}
\begin{aligned}
T_{re}&\approx  2.1\times10^{-3} \sqrt{(1-n_s)(11-3n_s)} \\
&\times \exp \left[-\frac{3 [2- 58 (1-n_s)] (w_{re}+1)}{(1-n_s) (3  w_{re}-1)}\right].
\end{aligned}
\end{equation}
Equations~ \eqref{ns:Nre} and \eqref{ns:Tre} are consistency relations that connect the reheating parameters ($T_{ re}$, $N_{ re}$, $w_{ re}$) to the inflationary observables ($n_s$, $r$) within the framework of non-minimally coupled Higgs inflation. Since the number of unknowns exceeds the number of independent observational constraints, the system is underdetermined: for each assumed value of $w_{ re}$, the data on $n_s$ determine a corresponding allowed range for $N_{ re}$ and $T_{ re}$. The resulting constraints are therefore model-dependent and should be interpreted as consistency bounds on the reheating parameters rather than as direct observational measurements. By varying the equation of state parameter $w_{re}$ during reheating, we can explore the impact of different reheating scenarios on the predictions of the Higgs inflation model with nonminimal coupling. 

To derive Eqs. \eqref{ns:Nre} and \eqref{ns:Tre}, the energy density at the end of inflation $\rho_e$ and the Hubble parameter at horizon crossing $H_*$ are computed under the slow-roll approximation rather than obtained from the exact numerical solution. 
To obtain the exact values of $\rho_e$, $H_*$, $N_*$, $n_s$, and $r$, we numerically solve both the background inflationary equations and the Mukhanov--Sasaki equations without assuming the strong-coupling limit. The numerical integration is performed using a fourth-order Runge--Kutta method, and the equations are evolved in conformal time with integration step
$
\Delta\tau={\varepsilon}/{(aH)},
$
where $\varepsilon\ll1$ is a control parameter. To test convergence, we repeat the calculation with smaller values of $\varepsilon$. Once the resulting observables, such as $n_s$ and $r$, become indistinguishable within numerical precision, the solution is regarded as converged and numerically stable.

For the boundary conditions, the field value at horizon crossing, $\phi_*$, is treated as a scanning parameter rather than being fixed a priori. Different choices of $\phi_*$ correspond to different horizon-crossing configurations and therefore lead to different values of $n_s$, $r$, and the $e$-folding number. To determine the corresponding field velocity $\dot{\phi}_*$ at horizon crossing, we make use of the attractor nature of slow-roll inflation. For a given $\phi_*$, we choose an initial field value before horizon crossing, $\phi_*+\Delta\phi$, and assign the initial velocity using the slow-roll approximation. We then numerically evolve the background equations until the field reaches $\phi_*$ and record the corresponding value of $\dot{\phi}_*$. This procedure is repeated by starting from earlier points, such as $\phi_*+2\Delta\phi$, $\phi_*+3\Delta\phi$, and so on. Once the values of $\dot{\phi}_*$ obtained at $\phi_*$ from two successive evolutions agree within numerical precision, we regard the solution as having reached the attractor trajectory and take this converged value as the field velocity at horizon crossing.

After $(\phi_*,\dot{\phi}_*)$ is determined in this way, the Hubble parameter at horizon crossing, $H_*$, is obtained directly from the background Friedmann equation. For both the scalar and tensor perturbations, we impose the Bunch--Davies initial condition deep inside the horizon,
$
u_k(\tau)={e^{-ik\tau}}/{\sqrt{2k}},
$
with $k|\tau|\gg1$ for a given $k$, and evolve the mode until it is well outside the horizon, where the corresponding curvature or tensor perturbation becomes constant. The end of inflation is defined by the condition
$
-{\dot H}/{H^2}=1.
$
Combining the  numerical results with Eqs. \eqref{Nre} and \eqref{eq:reh} allows us to evaluate the exact impact of reheating on the predictions of Higgs inflation with arbitrary $\xi$.

In the following section, we present our main results and demonstrate how the Higgs inflation model with nonminimal coupling can be brought into agreement with the latest P-ACT-LB constraints by taking reheating effects into account.

\section{Results}
\label{results}
\begin{figure}[H]
    \centering
    \includegraphics[width=1.0\linewidth]{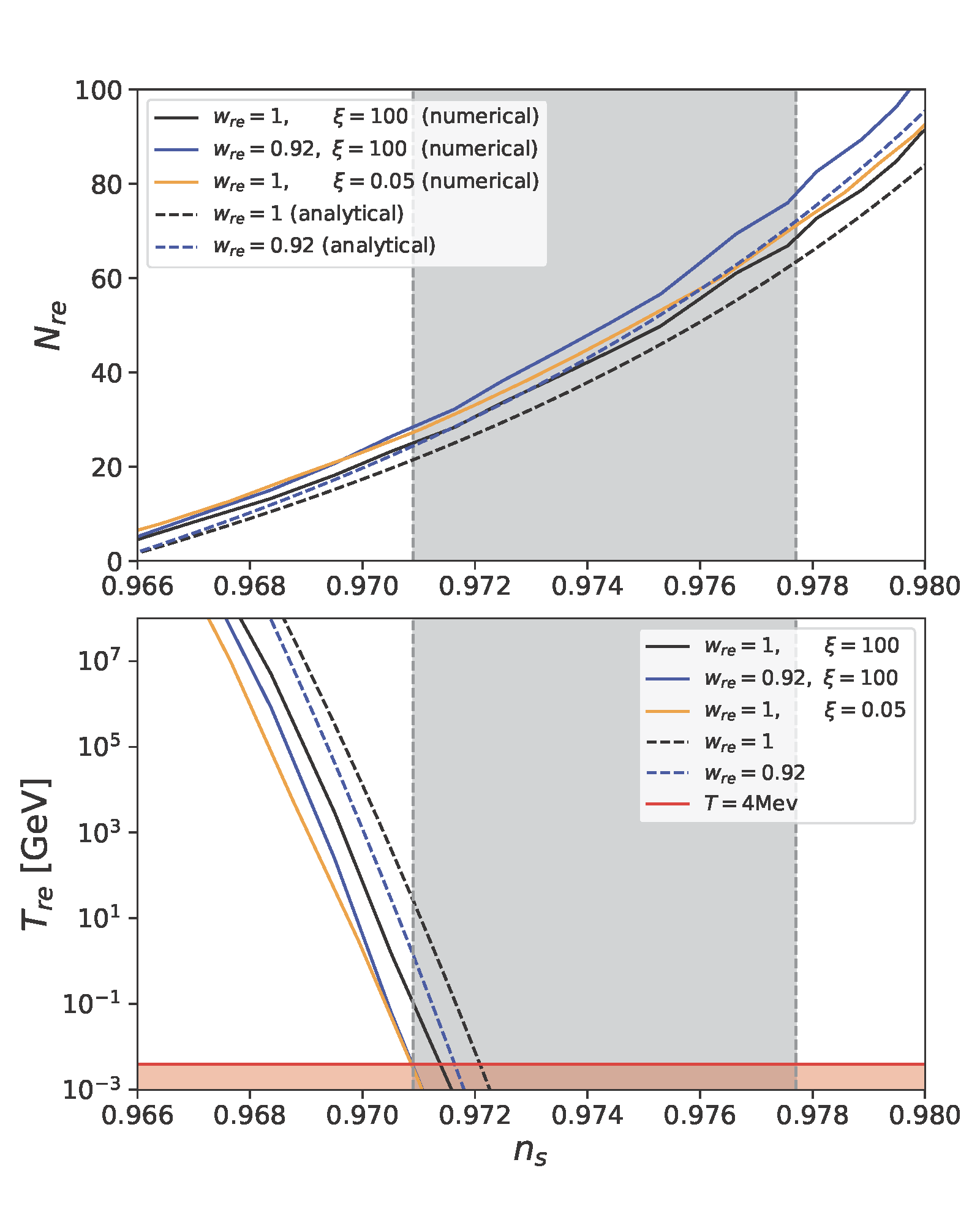}
    \caption{The relationship between the scalar spectral index $n_s$ and the reheating epoch.  
The upper panel shows how the number of  $e$-folds during reheating, $N_{re}$, affects $n_s$, while the lower panel displays the dependence of the reheating temperature $T_{re}$ on $n_s$.  Dashed curves correspond to results derived using the slow-roll approximations in Eqs. \eqref{ns:Nre} and \eqref{ns:Tre}, whereas solid curves represent results obtained via full numerical analysis.   The black and blue curves correspond to the cases with constant reheating equation of state parameters $w_{re} = 1$ and $w_{re} = 0.92$, respectively. 
 The orange curve shows the case with $w_{re} = 1$ and $\xi = 0.05$, which is the smallest value of $\xi$ that satisfies both the ACT and BBN constraints.    
 The vertical gray band indicates the 1$\sigma$ confidence interval from the P-ACT-LB data, $0.9743-0.0034 \leq n_s   \leq 0.9743+0.0034 $. The red horizontal line in the lower panel marks the lower bound on the reheating temperature from Big Bang Nucleosynthesis, $T_{re} \gtrsim 4\,\mathrm{MeV}$, below which scenarios are excluded~\cite{Kawasaki:1999na,Kawasaki:2000en,Hannestad:2004px,Hasegawa:2019jsa,Barbieri:2025moq}. 
}
 \label{fig:NTre}
\end{figure}

Using Eqs. \eqref{ns:Nre} and \eqref{ns:Tre}, we compute the effect of the reheating epoch on the scalar spectral index of the Higgs inflation model with nonminimal coupling under the slow-roll and strong-limit conditions, as illustrated in Fig. \ref{fig:NTre}. In addition, we numerically solve the background and perturbation equations of the Higgs model with a general nonminimal coupling $\xi$, and use Eqs. \eqref{Nre} and \eqref{eq:reh} to determine the exact impact of reheating. The corresponding results are also displayed in Fig. \ref{fig:NTre}.

The upper panel of Fig.\ref{fig:NTre} illustrates how the number of  $e$-folds during reheating, $N_{re}$, affects the scalar spectral index $n_s$.  The solid curves represent the numerical results, while the dashed curves correspond to the slow-roll approximation given by Eq.\eqref{ns:Nre}.   
As shown in the Fig.\ref{fig:NTre}, for the same reheating $e$-folding number $N_{re}$, the scalar spectral index predicted by the full numerical calculation differs from the slow-roll result by about $\Delta n_s \approx 0.001$. Conversely, for the same $n_s$, the corresponding difference in the reheating $e$-folding number $N_{re}$ between the numerical and slow-roll results is roughly $4$. 
Both discrepancies  mainly originate from the $\mathcal{O}(1)$ difference in the inflationary $e$-folding number $N_*$ between the slow-roll approximation and the full numerical result, together with neglected higher-order corrections in $1/N_*$. This difference of $N_*$ propagates into the predictions for $n_s$ and $N_{re}$, then into $T_{re}$, and finally into the inferred lower bound on $w_{re}$.

According to Eq. \eqref{N_Nre}, increasing the $e$-folding number $N_*$ requires the equation of state parameter during reheating to satisfy $w_{re} > 1/3$. As seen in Fig. \ref{fig:NTre}, the Higgs inflation model with nonminimal coupling can satisfy the 1$\sigma$ P-ACT-LB constraint on $n_s$ provided that the reheating duration is sufficiently long. However, Eq. \eqref{ns:Nre} also shows that the reheating temperature $T_{re}$ decreases exponentially with $N_{re}$. If $N_{re}$ becomes too large, the reheating temperature falls below the threshold required for successful Big Bang Nucleosynthesis (BBN), namely $T_{re} \gtrsim 4\,\text{MeV}$ \cite{Kawasaki:1999na,Kawasaki:2000en,Hannestad:2004px,Hasegawa:2019jsa,Barbieri:2025moq}, as illustrated in the lower panel of Fig. ~\ref{fig:NTre}.  While the slow-roll relations \eqref{ns:Nre} and \eqref{ns:Tre} qualitatively demonstrate how reheating influences inflationary predictions, precise constraints require full numerical analysis. Just as in the upper panel, the discrepancy between the analytical and numerical curves in the lower panel can also be understood as originating from the shift in the reheating $e$-folding number. Since $T_{re}\propto \exp[-3(1+w_{re})N_{re}/4]$, a shift of a few  $e$-folds in $N_{re}$ translates into a multiplicative offset in $T_{re}$, thereby explaining the difference between the slow-roll and numerical results. 
For the same $n_s$, the reheating temperature $T_{re}$ inferred from the numerical result is smaller than the slow-roll estimate by approximately a factor of $400$ for the curves shown in Fig.~\ref{fig:NTre}.

For the strong coupling case $\xi = 100$, numerical results indicate that reconciling both the P-ACT-LB data and BBN constraints requires the reheating equation of state to satisfy $w_{re} \geq 0.92$. Specifically, for $w_{re} = 0.92$, the scalar spectral index is $n_s =0.9743-0.0034$, the tensor-to-scalar ratio is $r=0.0024$, and the allowed value of $N_{re}$ is approximately $28.3$. In the case of $w_{re} = 1$, the P-ACT-LB constraint requires $24.9 \leq N_{re} \leq 68.4$, while the BBN condition restricts $N_{re} \leq 27.2$. Within this range, the predicted tensor-to-scalar ratio $r<0.0024$ remains consistent with observational bounds $r<0.036$ \cite{BICEP:2021xfz}. By contrast, the slow-roll approximation gives a weaker lower bound, $w_{re}\ge 0.81$. This difference originates from the fact that, for the same value of $n_s$, the reheating $e$-folding number obtained numerically is larger than that from the slow-roll approximation by roughly $4$ $e$-folds. As discussed above, this shift leads to a significantly lower reheating temperature. From Eq.~\eqref{ns:Tre}, for $w_{re}>1/3$ at fixed $n_s$, a lower reheating temperature implies a larger critical value of $w_{re}$.

Hence, while the slow-roll approximation is sufficiently accurate for the inflationary observables $n_s$ and $r$, it can induce non-negligible biases in the derived reheating quantities $N_{re}$, $T_{re}$, and the lower bound on $w_{re}$. This indicates that, in the high $w_{re}$ ($\geq 0.92$) and large $N_{re}$ ($24.9\text{--}28.3$) regime considered in this work, the slow-roll approximation remains reliable for inflationary observables, but should be used with greater caution when deriving reheating quantities. Although higher-order slow-roll corrections may improve the analytical approximation, they are not essential for the robustness of the present conclusions, since the final quantitative bounds in this work are extracted from the full numerical results.

For a general case without the strong coupling condition, we find that the smallest nonminimal coupling compatible with both the P-ACT-LB and BBN constraints is $\xi = 0.05$, assuming $w_{re} = 1$. This scenario yields a scalar spectral index at the lower edge of the 1 $\sigma$ region, $n_s = 0.9743 - 0.0034$, a tensor-to-scalar ratio of $r = 0.0092$, an $e$-folding during reheating of $N_{re} = 27.2$, and a reheating temperature at the end of reheating  of $T_{re} = 4 \rm {MeV}$. For $\xi < 0.05$, achieving consistency with P-ACT-LB data would require either $w_{re} > 1$, which is physically disfavored, or a reheating temperature below the BBN threshold, which is excluded.

The above results demonstrate that by considering the reheating epoch, the Higgs inflation model with nonminimal coupling can be reconciled with the latest P-ACT-LB constraints. The reheating phase plays a crucial role in determining the inflationary observables, and by choosing an appropriate equation of state during reheating, the model can satisfy both the P-ACT-LB data and the temperature requirement for successful BBN.

Furthermore, the universal attractor relation $n_s -1 = -2/N_*$ remains consistent with the latest observational data when the effects of reheating are taken into account. This implies that not only the Higgs inflation model with nonminimal coupling but also the $R^2$ inflation model, which shares the same attractor relation, can be made compatible with the P-ACT-LB constraints by considering the reheating phase.

While our analysis demonstrates that an equation of state parameter $w_{re} \geq 0.92$ during reheating can reconcile Higgs inflation model with nonminimal coupling with ACT DR6 data, it is important to examine what physical mechanisms could produce such values. The standard reheating scenario, where the inflaton oscillates in a quadratic potential, yields $w_{ re} = 0$ and is insufficient. 
A natural way to obtain a large reheating equation of state is to consider a kination–dominated phase, during which the kinetic energy of the inflaton overwhelms its potential energy~\cite{Spokoiny:1993kt,Ferreira:1997hj,Peebles:1998qn}.
 
For the Higgs inflation model with nonminimal coupling specifically, a reheating equation of state $w_{re}\ge0.92$ can be accommodated by an effective modification of the potential in the small-field regime.  Motivated by quintessential inflation~\cite{Spokoiny:1993kt,Ferreira:1997hj}, to obtain a kination domination, we adopt a simple exponential tail in the Einstein frame \cite{Sahni:2001qp},
\begin{equation}\label{eq:U_eff}
U_{\rm kin}(\psi)= U_e \exp \big[-\lambda_{\rm kin}(\psi_e-\psi)\big], \quad \psi<\psi_e,
\end{equation}
where $U_e \equiv U(\psi_e)$ ensures continuity and $\lambda_{\rm kin}\gg 1$ controls the steepness of the small-field tail. Once the inflaton crosses into the exponential region, $\psi<\psi_e$, the slope becomes large,  and the field rapidly accelerates such that the kinetic energy dominates over the
potential energy $\frac12 \dot\psi^2 \gg U(\psi)$.
The inflaton then behaves as a stiff fluid with equation of state $w\simeq 1$,
and its energy density redshifts as
\begin{equation}
\rho_\psi \simeq \frac12\dot\psi^2 \propto a^{-6},
\label{eq:kin_rho}
\end{equation}
realizing a kination phase. Meanwhile, any subdominant radiation component produced at the end of inflation, for
example through gravitational particle production~\cite{Ford:1986sy,Kolb:2023ydq}
or via a weakly coupled spectator field \cite{Felder:1998vq,Dimopoulos:2018wfg}, redshifts as $\rho_r\propto a^{-4}$, and eventually overtakes $\rho_\psi$, thereby completing reheating and initiating the standard radiation-dominated phase. We emphasize that this construction is phenomenological and does not assume any particular UV completion for the small-field regime.

\section{Conclusion and Discussion}
\label{Conclusion}
Recent observational data from the Atacama Cosmology Telescope have established new constraints on the scalar spectral index $n_s$.   A joint analysis of the results from ACT, {\it Planck}, and DESI (P-ACT-LB) has yielded the constraint  $n_s = 0.9743\pm 0.0034$. This observational data  disfavor the Higgs inflation model with a nonminimal coupling term, in which 
the scalar spectral index  is $n_s-1 \approx   - 2/N_*$, under the strong coupling condition. This formula is a  universal attractor for a wide range of inflationary models, such as hilltop inflation, T model, E model,  the Starobinsky model. The tension between the Higgs inflation model with nonminimal coupling  and the P-ACT-LB constraints primarily stems from the conventional assumption that the $e$-folding number at horizon crossing before the end of inflation occurs within the range $N_* = 50-60$.

In this paper, we have investigated the impact of the reheating epoch on the predictions of the Higgs inflation model with nonminimal coupling in light of the latest observational constraints from the P-ACT-LB data. We have shown that by considering the effects of reheating, the Higgs inflation model with nonminimal coupling  can be reconciled with the P-ACT-LB constraints, which favor a higher value of the scalar spectral index $n_s$ compared to the canonical predictions of the model.

Our analysis demonstrates that the duration of the reheating phase,  $e$-folding number $N_{re}$, plays a crucial role in determining the inflationary observables. By varying the equation of state parameter $w_{re}$ during reheating, we have explored the impact of different reheating scenarios on the predictions of the Higgs inflation model with nonminimal coupling, and the results are obtained by the numerical method.  
For strong coupling case with $\xi=100$, we have found that for $w_{re} \geq 0.92$, the model can simultaneously satisfy the 1$\sigma$ constraint from the P-ACT-LB data and the temperature requirement for successful BBN. Specifically, when $w_{re}=0.92$, the $e$-folding number during reheating is $N_{re} =28.3$, when $w_{re}=1$, it must satisfy $24.9 \leq N_{re} \leq 27.2$.  For a general case without the strong coupling condition, we find that the smallest nonminimal coupling parameter compatible with both the P-ACT-LB and BBN constraints is $\xi= 0.05$ with $w_{re}=1$, and the $e$-folding number during reheating is $N_{re} = 27.2$.

Our results emphasize the importance of considering the reheating phase in inflationary model building and the need for a more comprehensive analysis of the post-inflationary dynamics. The reheating epoch serves as a bridge between the inflationary phase and the subsequent thermal history of the universe, and its properties can significantly influence the inflationary predictions.

In addition to observational constraints, theoretical developments in the understanding of the reheating process  will be crucial for refining the predictions of inflationary models. Different reheating scenarios, including the standard or non-oscillatory mechanisms, can lead to distinct evolutions of the inflaton and radiation energy densities, thereby modifying the $e$-folding number and the associated predictions for $(n_s,r)$. A more detailed understanding of these possibilities may reveal new avenues for reconciling inflationary models with precision observations.

It is important to note that achieving the required equation of state $w_{ re} \geq 0.92$ poses additional model-building challenges. Standard reheating mechanisms typically yield $w_{  re} \leq 1/3$, and obtaining values close to unity requires specific modifications to the inflationary model. For instance, the Starobinsky $R^2$ model, which shares the same attractor behavior as Higgs inflation model with nonminimal coupling, predicts $w_{ re} = 0$ in its standard formulation \cite{Bassett:2005xm}. Reconciling such models with ACT DR6 data through reheating thus requires additional theoretical ingredients beyond their original elegant formulations. This suggests that while our mechanism provides a possible resolution to the tension with ACT  DR6 data, it comes at the cost of increased model complexity. Future work should explore concrete realizations of such reheating scenarios and assess their naturalness within specific inflationary frameworks.

In conclusion, our work demonstrates that the Higgs inflation model with nonminimal coupling remains a viable candidate for describing the early universe, even in the face of the latest observational constraints from P-ACT-LB. By taking into account the effects of reheating, we have shown that the model can be reconciled with the data, providing a compelling framework for understanding the origin of the universe's structure and evolution. As future observations and theoretical developments continue to shape our understanding of inflation and the reheating epoch, the Higgs inflation model with nonminimal coupling will undoubtedly remain a subject of active research and exploration.

Note added: While finalizing this work, we found two parallel independent studies \cite{Drees:2025ngb,Zharov:2025evb} that also consider the reheating phase to explain the ACT DR6 observational data.

\Acknowledgements{
We thank  the insightful correspondence with Jun'ichi Yokoyama. 
LL is supported by the National Natural Science Foundation of China (Grant No.~12505054 and 12447101) and the Fundamental Research Funds for the Central Universities.
This work is supported in part by the National Natural Science Foundation of China under Grant No. 12205015,  No. 12433001, No. 12175184, No. 12305075, No. 12588101, No. 12535002, and the National Key Research and Development Program of China under Grant No. 2023YFC2206704.
\\
\\
Conflict of Interest: The authors declare that they have no conflict of interest.
}

\bibliographystyle{scpma}
\bibliography{ref}

\end{multicols}
 
\end{document}